\documentclass{article}
\usepackage{emsnewsletter}

\usepackage{todonotes}

\usepackage{amssymb,amsmath,amsfonts,amsthm}

\usepackage{xcolor}
\usepackage{graphicx}

\usepackage[framemethod=TikZ]{mdframed}
\usepackage{lipsum}
\mdfdefinestyle{MyFrame}{%
    linecolor=black,
    outerlinewidth=2pt,
    roundcorner=20pt,
    innertopmargin=\baselineskip,
    innerbottommargin=\baselineskip,
    innerrightmargin=20pt,
    innerleftmargin=20pt,
    backgroundcolor=white}

\DeclareMathOperator{\Span}{span}
\newcommand{\bra}[1]{\langle#1|}
\newcommand{\ket}[1]{|#1\rangle}
\newcommand{\braket}[2]{\langle#1|#2\rangle}

\newcommand{\eye}{\mathbf{1}}

\def\ketbra#1#2{{\vert#1\rangle\!\langle#2\vert}}

\title{TENSOR NETWORKS IN MACHINE LEARNING} 

\author{Richik Sengupta, Soumik Adhikary, Ivan Oseledets and Jacob Biamonte \\
Skolkovo Institute of Science and Technology, Nobel Street 3, Skolkovo 143025, Russian Federation.}

\begin{document} 
\maketitle 

\begin{abstract}

A tensor network is a type of decomposition used to express and approximate large arrays of data.  A given data-set, quantum state or higher dimensional multi-linear map is factored and approximated by a composition of smaller multi-linear maps.  This is reminiscent to how a Boolean function might be decomposed into a gate array: this represents a special case of tensor decomposition, in which the tensor entries are replaced by 0, 1 and the factorisation becomes exact.  The collection of associated techniques are called, tensor network methods: the subject developed independently in several distinct fields of study, which have more recently become interrelated through the language of tensor networks.  The tantamount questions in the field relate to expressability of tensor networks and the reduction of computational overheads.  A merger of tensor networks with machine learning is natural. On the one hand, machine learning can aid in determining a factorization of a tensor network approximating a data set.  On the other hand, a given tensor network structure can be viewed as a machine learning model.  Herein the tensor network parameters are adjusted to learn or classify a data-set.  In this survey we recover the basics of tensor networks and explain the ongoing effort to develop the theory of tensor networks in machine learning.   

\end{abstract}

\section{Introduction}

Tensors networks are ubiquitous in most areas of modern science including  data science \cite{Cic14}, condensed matter physics \cite{Gon+15}, string theory \cite{JE21} and quantum computer science. The manner in which tensors are employed/treated exhibit significant overlap across many of these areas. In data science tensors are used to represent large datasets. In condensed matter physics and in quantum computer science, tensors are used to represent states of quantum systems. 

Manipulating large tensors comes at a high computational cost \cite{Nov+15}. This observation has inspired techniques for tensor decompositions that would reduce computational complexity while preserving the original data that it represents. A collection of such techniques are now known as tensor network methods.

Tensor networks have arose to prominence in the last fifteen years with several European schools playing leading roles in their modern development, as a means to describe and approximate quantum states (see the review \cite{Cir+21}). The topic however dates back much further,  to the works of Penrose~\cite{Pen71} and in retrospect, even arose as  special cases in the works of Cayley~\cite{Cay54}.  Tensor networks have a deep modern history in mathematical physics \cite{Pen71}, in category theory \cite{Sel11}, in computer science, algebraic logic and related disciplines \cite{BS11}.  Such techniques are now becoming more common in data-science and machine learning (see the reviews \cite{Cic+16, Cic+17}).

\subsection{Rudimentary ideas about multi-linear maps}
 It might be stated that the objective of linear algebra is to classify linear operators upto isomorphism and study the simplest representative in each equivalence class. This motivates the prevalence of decompositions including SVD, LU and the Jordan normal form. A special case of linear operators are linear maps from a vector space $V$ to an arbitrary field like $\mathbb{C}$ or $\mathbb{R}$. The set of linear maps form the dual space (vector space of covectors) $V^\star$ to our vector space $V$. 

A natural generalization of linear maps are multilinear maps which are linear in each argument when the values of other arguments are fixed.
For a given pair of integers $p$ and $q,$ a type $(p, q)$ tensor T is defined as a multilinear map,
\begin{equation}\label{tm}
      T:\underbrace {V^{*}\times \dots \times V^{*} } _{p{\text{ copies}}}\times \underbrace {V\times \dots \times V} _{q{\text{ copies}}}\rightarrow \mathbb{K}
\end{equation}
where $\mathbb{K}$ is an arbitrary field. The tensor $T$ is an order (valence) $p+q$ tensor. Note that some authors refer to this as rank  $p+q$ but we will never do that.

It is often more convenient to view tensors as elements of a vector space known as the tensor product space. Thus, the above $(p, q)$ tensor $T$ in this new paradigm can  be defined as
\begin{equation}
    T\in \underbrace {V\otimes \dots \otimes V} _{p{\text{ copies}}}\otimes \underbrace {V^{*}\otimes \dots \otimes V^{*}} _{q{\text{ copies}}}.
\end{equation}

Moreover, the universal property of tensor products of vector spaces states that any multilinear map can always be replaced by a unique linear map acting from the tensor product of vector spaces to the base field.

If we assume the axiom of choice,  any vector space admits a Hamel basis. The components of a tensor $T$ are therefore the coefficients of the tensor with respect to the basis ${\bf e}_i$ of V and its dual basis $\boldsymbol{\epsilon}^j$(basis of the dual space $V^{*}$), that is
\begin{equation}\label{tps}
    T=T_{j_{1}\dots j_{q}}^{i_{1}\dots i_{p}}\;\mathbf {e} _{i_{1}}\otimes \cdots \otimes \mathbf {e} _{i_{p}}\otimes {\boldsymbol {\epsilon }}^{j_{1}}\otimes \cdots \otimes {\boldsymbol {\epsilon }}^{j_{q}}.
\end{equation}
Summation over repeating indices is implied in \eqref{tps} as per the Einstein summation convention.

Returning to (\ref{tm}), given a tuple of $p$ covectors  $(c^{1}, \ldots,c^{p})$ and $q$ vectors $(v_1,\ldots, v_q)$ the value of the tensor is evaluated as

\begin{equation}\label{teval}
    T_{j_{1}\dots j_{q}}^{i_{1}\dots i_{p}}\;c^1(\mathbf {e} _{i_{1}})\times \cdots \times c^p(\mathbf {e} _{i_{p}})\times {\boldsymbol {\epsilon }}^{j_{1}}(v_1)\times \cdots \times {\boldsymbol {\epsilon }}^{j_{q}}(v_q) \in \mathbb{K}.
\end{equation}

Where $c^k(\mathbf {e}_{i_k})$ and ${\boldsymbol{\epsilon}}^{j_l}(v_l)$ are numbers obtained by evaluating covectors (functionals) at the corresponding vectors.
In quantum computation, the basis vectors ${\bf e}_{i_k}$ are denoted by $\ket{i_k}$ and the basis covectors ${\boldsymbol {\epsilon }}^{j_{l}}$ are denoted by $\bra{j_l}$.
The tensor products are succinctly written in Dirac notation as:
\begin{equation}
  \mathbf {e} _{i_{m}} \otimes \mathbf {e} _{i_{l}} := \ket{i_m i_l}; \quad {\boldsymbol {\epsilon }}^{j_{m}}\otimes {\boldsymbol {\epsilon }}^{j_{l}}:=\bra{j_m j_l}; \quad \mathbf {e} _{i_{m}}\otimes {\boldsymbol {\epsilon }}^{j_{l}} := \ketbra{i_{m}}{j_{l}}
\end{equation}

And the tensor $T$ takes the form

\begin{equation}
    T=T_{j_{1}\dots j_{q}}^{i_{1}\dots i_{p}}\;\ketbra{i_1\ldots i_p}{j_1\ldots j_p}.
\end{equation}

Similarly, given a tuple of $p$ covectors  $(c^{1},\ \ldots,c^{p})$ and $q$ vectors $(v_1,\ldots, v_q)$ we write them as  $\bra{c_{1} \ldots c_{p}}$ and $\ket{v_1 \ldots v_q},$ as elements of the corresponding tensor product space(s).

Now, the evaluation (\ref{teval}) takes the form:

\begin{equation}
    T_{j_{1}\dots j_{q}}^{i_{1}\dots i_{p}}\;\braket{c_{1} \ldots c_{p}}{i_1\ldots i_p}\braket{j_1\ldots j_p}{v_1 \ldots v_q}.
\end{equation}

From Riesz's theorem on linear functionals, it follows that the evaluation $\braket{*}{*}$ above can be seen as an inner product owing to the fact that in quantum computation the vector spaces in consideration are Hilbert spaces and their duals.

Finally, a tensor can be identified with the array of coefficients $T_{j_{1}\dots j_{q}}^{i_{1}\dots i_{p}}$ in a specific basis decomposition. This approach is not basis independent but useful in applications. Henceforth in this review we will fix the standard basis which will establish a canonical isomorphism between the vector space and its dual.  This, for example in the simplest case of $p = q = 1$, gives us the following  equivalence:

\begin{equation}
    T_{ij} \cong T_{ji} \cong T^{ij} \cong T^{ji} \cong T^{i}_{j} \cong T^{j}_{i}
\end{equation}

In the more general case this leads to the equivalence between $T_{j_{1}\dots j_{q}}^{i_{1}\dots i_{p}}$ and $T_{j_{1}\dots j_{q} i_{1}\dots i_{p}}$. Given a valence $m$ tensor $T$, the total number of tensors in the equivalency class formed by raising, lower and exchanging indices has cardinality $(1+m)!$ \cite{BB17}. Recognizing this arbitrariness, Penrose considered a graphical depiction of tensors \cite{Pen71}, stating that, ``it now ceases to be important to maintain a distinction between upper and lower indices.”  This is the convention readily adopted in contemporary literature.

\subsection{Tensor trains a.k.a.~matrix product states}
\label{sec:ten_net}

Consider a tensor $T$ with components $T_{j_1 j_2 \cdots j_n}$ with $j_k = 1, 2,$ $ \cdots, d$. $T$ has $d^n$ components, a number that can exceed the total number of electrons in the universe with $d$ as small as $2$ and $n \approx 300$. Clearly storing the components of such a large tensor in a memory and subsequently manipulating it can become impossible. Nevertheless for most practical purposes a tensor typically contains a large amount of redundant information. This enables factoring of $T$ into a sequence of ``smaller" tensors.

Tensor trains \cite{Ose11} and matrix product states (MPS) \cite{Gar+07,Oru14} arose in data science and in condensed matter physics, where it was shown that any tensor $T$ with components $T_{j_1 j_2 \cdots j_n}$ admits a decomposition of the form:

\begin{equation} \label{eq:ttobc}
    T_{j_1 j_2 \cdots j_n} = \boldsymbol{\alpha}^\dagger A_{j_1}^{(1)} A_{j_2}^{(2)} \cdots A_{j_n}^{(n)} \boldsymbol{\beta}
\end{equation}
where $A_{j_k}^{(k)}$ is a $r_{k-1} \times r_{k}$ dimensional matrix with $\boldsymbol{\alpha}$ and $\boldsymbol{\beta}$ as $r_0$ and $r_n$ dimensional vectors respectively. 

\begin{figure}[!ht] 
    \centering
    \includegraphics[width=0.5\textwidth]{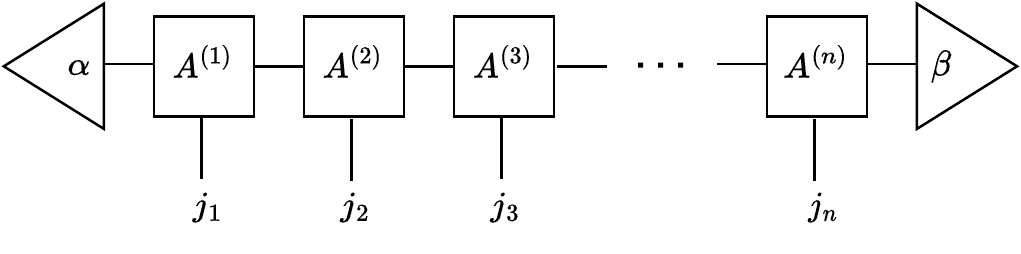}
    \caption{Graphical representation of tensor trains (open boundary condition - matrix product state representation). See \eqref{eq:ttobc}, \eqref{eq:mpsobc}.
    Each index in the tensor $T_{j_1 \cdots j_n}$ is represented in the diagrams by an open wire, pointing downwards. We call them the physical bonds. The horizontal wires, represent extra indices which are summed over. Such internal wires are known as the virtual bonds.} \label{fig:mpsobc}
\end{figure}

Likewise a $n$-qubit state $\ket{\psi} \in (\mathbb{CP}^{2})^{\otimes n}$  written in the computational basis as $\ket{\psi} = \sum_{j_1 j_2 \cdots j_n} T_{j_1 j_2 \cdots j_n} \ket{j_1 j_2 \cdots j_n}$, $j_k \in \{0,1\}$ can equivalently be expressed as:

\begin{equation} \label{eq:mpsobc}
    \ket{\psi} = \sum_{j_1 \cdots j_n} \bra{\alpha}A^{(1)}_{j_1} A^{(2)}_{j_2} \cdots A^{(n)}_{j_n} \ket{\beta} \ket{j_1 j_2 \cdots j_n}.
\end{equation}
Here $A_{j_k}^{(k)}$ is a $r_{k-1} \times r_{k}$ dimensional matrix and $\ket{\alpha}, \ket{\beta}$ are $r_0$ and $r_n$ dimensional vectors. Note that here we are adhering to the braket notation, as is customary in quantum mechanics. The representation of $\ket{\psi}$ in \eqref{eq:mpsobc} is called the matrix product state representation with an open boundary condition (OBC-MPS). See figure~\ref{fig:mpsobc} for the graphical representation.

\begin{figure}[!ht] 
    \centering
    \includegraphics[width=0.43\textwidth]{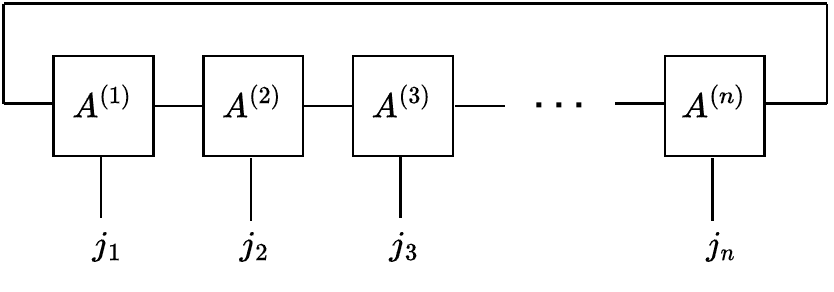}
    \caption{Graphical representation of the matrix product state in \eqref{eq:mpspbc}}.\label{fig:mpspbc}
\end{figure}

Yet another useful MPS decomposition that a state might admit is the MPS with periodic boundary condition (PBC-MPS) \cite{Oru14}. The PBC-MPS representation of a $n$-qubit state $\ket{\psi} = \sum_{j_1 j_2 \cdots j_n} T_{j_1 j_2 \cdots j_n} \ket{j_1 j_2 \cdots j_n}$, $j_k = 0,1$ is given by:

\begin{equation} \label{eq:mpspbc}
     \ket{\psi} = \sum_{j_1...j_n} Tr(A^{(1)}_{j_1} A^{(2)}_{j_2} \cdots A^{(n)}_{j_n}) \ket{j_1 j_2 ... j_n}
\end{equation}
where $A^{(k)}_{j_k}$ is a $r \times r$ matrix. The graphical representation of a PBC-MPS is shown in figure~\ref{fig:mpspbc}.

A $n$ qubit state $\ket{\psi} = \sum_{j_1 \cdots j_n} T_{j_1...j_n} \ket{j_1 ... j_n}$ has $2^n$  independent coefficients $T_{j_1 ... j_n}$. The MPS representation of $\ket{\psi}$ on the other hand is less data intensive. If $A^{(k)}_{j_k}$ is a $r \times r$ matrix for all $k$, the size of the representation becomes $2n r^2$, which is linear in $n$ for a constant $r$. The point of the method is to choose $r$ such that a good and compact approximation of $\ket{\psi}$ is obtained. $r$ is often also called as the virtual bond dimension. Data compression becomes even more effective if the MPS is site independent, that is if $A^{(k)}_{j_k} = A_{j_k} \forall k$. It has been shown that a site independent representation of a PBC-MPS always exists if the state is translationally invariant \cite{Gar+07}. Note that MPS is invariant under the transformation $A_j \rightarrow P A_j P^{-1}$ for any invertible $P$, which follows from  the cyclicity of the trace operator. Therefore, it is often customary to impose an additional constraint here viz. $\sum_j A_j A_j^\dagger = \eye$ in order to fix the gauge freedom \cite{Cir+21} (see also the connections to algebraic invariant theory \cite{BBL13}).

\section{Machine learning: classical to quantum}

\subsection{Classical machine learning}

At the core of machine learning is the task of data classification. In classification, we are typically provided with a labelled dataset $\mathcal{S} = \{(\mathbf{x}_j, \mathbf{y}_j)\}_{j=1}^M$ where ${\mathbf{x}_j} \in \mathbb{R}^N$ is the input data (e.g.~animal images) and $\mathbf{y}_j \in \mathbb{R}^{d}$ is the corresponding label (e.g.~type of animal). The objective is to find a suitable machine learning model $F$ with tunable parameters ${\boldsymbol{\theta}} \in \mathbb{R}^k$ that generates the correct label for a given input $\mathbf{x} \in \mathbb{R}^N$. Note that our model can be looked upon as a  family of functions parametrized by ${\boldsymbol{\theta}}.$ $F$ takes a data vector as an input and outputs a predicted label; for an input datum ${\bf x}_j$ the predicted label is $F(\mathbf{x}_j, {\boldsymbol{\theta}})$. To ensure that our model generates the correct labels, it needs to be trained; in order to accomplish this, a training set $\mathcal{T} \subset \mathcal{S}$ is chosen, elements of which serve as input data to train $F$.  Training requires a cost function:

\begin{equation}
\label{eq:loss_fn_formal}
    \mathcal{L} ({\boldsymbol{\theta}}) = \sum_{(x_j,y_j)\in \mathcal{T}} D(F(\mathbf{x}_j, {\boldsymbol{\theta}}), \mathbf{y}_j)
\end{equation}
where $D(\cdot, \cdot)$ estimates the mismatch between the real label and the estimated label. Typical choices for $D$ includes e.g.~the negative log-likelihood function, mean squared errors (MSE) etc.~\cite{ros04}.  By minimizing \eqref{eq:loss_fn_formal} with respect to ${\boldsymbol{\theta}},$  ${ \boldsymbol{\theta}}^\star \in \arg \min_{\boldsymbol{\theta}}$ $ \mathcal{L} ({\boldsymbol{\theta}})$ is found which completes the training. After the the model $F$ has been trained we can evaluate its performance by feeding it inputs from $\mathcal{S} \setminus \mathcal{T}$ (often referred to as the validation set) by checking for classification accuracy. 
 
For a more formal description, let us assume that the dataset $\mathcal{S}$ is sampled from a joint probability distribution with a density function $p(\mathbf{x}, \mathbf{y})$. The role of a classifier model is to approximate the conditional distribution $p(\mathbf{y} \vert \mathbf{x})$. The knowledge of $p(\mathbf{x}, \mathbf{y})$ allows us, in principle, to establish theoretical bounds on the performance of the classifier. Consider the generalisation error (also called risk) 
defined as $ \mathcal{G} ({\boldsymbol{\theta}}) = \mathbb{E}_{p(\mathbf{x}, \mathbf{y})} (D(F(\mathbf{x}, {\boldsymbol{\theta}}),\mathbf{y})).$ A learning algorithm is said to generalise if $\lim_{M^\prime \rightarrow \infty} \frac{\mathcal{L} ({\boldsymbol{\theta}})}{M^\prime} = \mathcal{G} ({\boldsymbol{\theta}}),$  here $M^\prime$ is the cardinality of the training set. However, since in general we do not have access to $p (\mathbf{x},\mathbf{y})$ we can only attempt to provide necessary conditions to bound the difference of the generalization error and the empirical error by checking certain stability conditions to ensure that our learning model is not too sensitive to noise in the data \cite{BE02}. For example we can try to ensure that our learning model is not affected if one of the data points is left out during training. The technique of regularization prevents overfitting.

Several different types of machine learning models $F$ exist, which ranges from fairly elementary models such as perceptrons to highly involved ones such as deep neural networks (DNN) \cite{LBH15, Ros58}.  The choice of $F$ is heavily dependent on the classification task, type of the dataset and the desired training properties. Consider a dataset $\mathcal{S}$ with two classes (a binary dataset) that is linearly separable. That is (i) $\mathbf{y}_j \in \{-1,1\}$ and (ii) it is possible to construct a hyperplane which separates the input data belonging to the different classes. Finding this hyperplane a.k.a.~the decision boundary is therefore sufficient for data classification in $\mathcal{S}$. This task can be accomplished with a simplistic machine learning model---the perceptron---which is in fact a single McCulloch-Pitts neuron \cite{mccul43}.  The algorithm starts with the candidate solution for the hyperplane $({\bf w}^\top\cdot{\bf x} + b) = 0$ where ${\bf w}, b$ are tunable parameters and play the role of $\boldsymbol{\theta}$. It is known from the perceptron convergence algorithm, that one can always find a set of parameters ${\bf w} = {\bf w}^\star, b = b^\star$, such that $\forall {\bf x}_j : \mathbf{y}_j = -1$, $({\bf w}^\star{^\top}\cdot{\bf x}_j + b^\star) \leq 0$ while $\forall {\bf x}_j : \mathbf{y}_j = 1$, $({\bf w}^\star{^\top}\cdot{\bf x}_j + b^\star) > 0$.

Most datasets of practical importance however are not linearly separable and as a result cannot be classified by the perceptron model by itself. Assuming that $\mathcal{S}$ is a binary dataset which is not linearly separable, we consider a map $\Lambda: \mathbb{R}^N \rightarrow \Gamma$, $dim( \Gamma)> N$, with the proviso that $\Lambda$ is non-linear in the components of its inputs \cite{bishop06}.  In machine learning $\Lambda$ is called a feature map while the vector space $\Gamma$ is known as a feature space. Thus $\Lambda$ can non-linearly map an input datum ${\bf x}_j$ to a vector $\Lambda({\bf x}_j)$ in the feature space. The significance of this step follows from Cover's theorem on separability of patterns \cite{cov65} which suggests that the transformed dataset $\mathcal{S}^\prime = \{(\Lambda({\bf x}_j), \mathbf{y}_j)\}_{j=1}^M$ is more likely to be linearly separable. For a good choice of $\Lambda$, the data classification step now becomes straightforward as it would now be enough to fit a hyperplane to separate the two classes in the feature space. Indeed the hyperplane can be constructed, using the perceptron model, provided the feature map $\Lambda$ is explicitly known. Nevertheless a hyperplane can still be constructed, if $\Lambda$ is not explicitly known. A particularly elegant way to this is via the support vector machine (SVM) \cite{cort95},  by employing the so called Kernel trick \cite{shawe04}. 

Yet again we consider the binary dataset $\mathcal{S}^\prime = \{(\Lambda({\bf x}_j), \mathbf{y}_j)\}_{j=1}^M$. The aim is to construct a hyperplane to separate the samples  belonging to the two classes. In addition we would like to maximize the margin of separation. Formally we search for a set of parameters $\mathbf{w} = \mathbf{w}^\star, b = b^\star$ such that $\forall {\bf x}_j : \mathbf{y}_j = -1$, $({\bf w}^\star{^\top}\cdot\Lambda({\bf x}_j) + b^\star) \leq -1$ and  $\forall {\bf x}_j : \mathbf{y}_j = 1$, $({\bf w}^\star{^\top}\cdot\Lambda({\bf x}_j) + b^\star) \geq 1$. As in the perceptron model the algorithm in SVM starts with the candidate solution $({\bf w}^\top\cdot{\bf x} + b)$ and the parameters ${\bf w}, b$ are tuned based on the training data. An interesting aspect of the SVM approach however, is the dependence of the algorithm on a special subset of the training data called support vectors, the ones that satisfies the relation $({\bf w}^\star{^\top}\cdot\Lambda({\bf x}_j) + b^\star) = \pm 1$. We assume that there are $S$ such vectors $\Lambda({\bf x}_j^{(s)})$. With some algebra, which we would omit here, it can be shown that the decision boundary is given by the hyperplane:

\begin{equation}
    \label{eq:hyperplane_svd}
    \sum_{j=1}^S \alpha_j^\star \mathbf{y}_j [\Lambda({\bf x}_j^{(s)})^\top \Lambda({\bf x})] + b^\star = 0
\end{equation}
where,

\begin{equation}
    \label{eq:hyperplane_svd_b}
    b^\star = \frac{1}{S} \sum_{j=1}^S \Big( \mathbf{y}_j - \sum_{k=1}^S \alpha^\star_j \mathbf{y}_j [\Lambda({\bf x}_j^{(s)})^\top \Lambda({\bf x}_k^{(s)})] \Big)
\end{equation}
and

\begin{equation}
\label{eq:alpha_opt}
  {\boldsymbol{\alpha}^\star} = \arg \max_{\boldsymbol{\alpha}} \Big(\sum_j \alpha_j - \frac{1}{2}\sum_j \sum_k \alpha_j \alpha_k \mathbf{y}_j \mathbf{y}_k \Lambda({\bf x}_j)^\top \Lambda({\bf x}_k) \Big)
\end{equation}
with additional conditions: $\sum_j \alpha_j \mathbf{y}_j=0$ and $\alpha_j \geq 0$. All summations in \eqref{eq:alpha_opt} are over the entire training set. We make a key observation here: from \eqref{eq:hyperplane_svd}, \eqref{eq:hyperplane_svd_b} and  \eqref{eq:alpha_opt} we see that the expression of the decision boundary has no explicit dependence on the feature vectors $\Lambda({\bf x}_j)$. Instead the dependence is solely on the inner product of type $\Lambda({\bf x}_j)^\top \Lambda({\bf x}_k)$. This allows us to use the kernel trick. 

\begin{mdframed}[style=MyFrame]
\subsection*{Support functions in optimisation}

The support function $ h_{C}\colon \mathbb {R} ^{n}\to \mathbb {R} $ of a non-empty closed convex set $C \subset  \mathbb {R} ^{n}$ is given by\linebreak
    $ h_{C}(x)=\sup\{\braket{x}{c}: c\in C, x\in {\mathbb {R}}^{n}\}.$
The hyperplane $H(x)=\{y\in {\mathbb {R}}^{n}:\braket{y}{x}=h_{C}(x)\}$ is called a supporting hyperplane of $C$ with exterior unit normal vector $x.$
The function $h_C(x)$ outputs the signed distance of $H(x)$ from the origin.
The data points that lie on $H(x)$ are called support vectors in the machine learning literature.
\begin{center}
    \includegraphics[width=\textwidth]{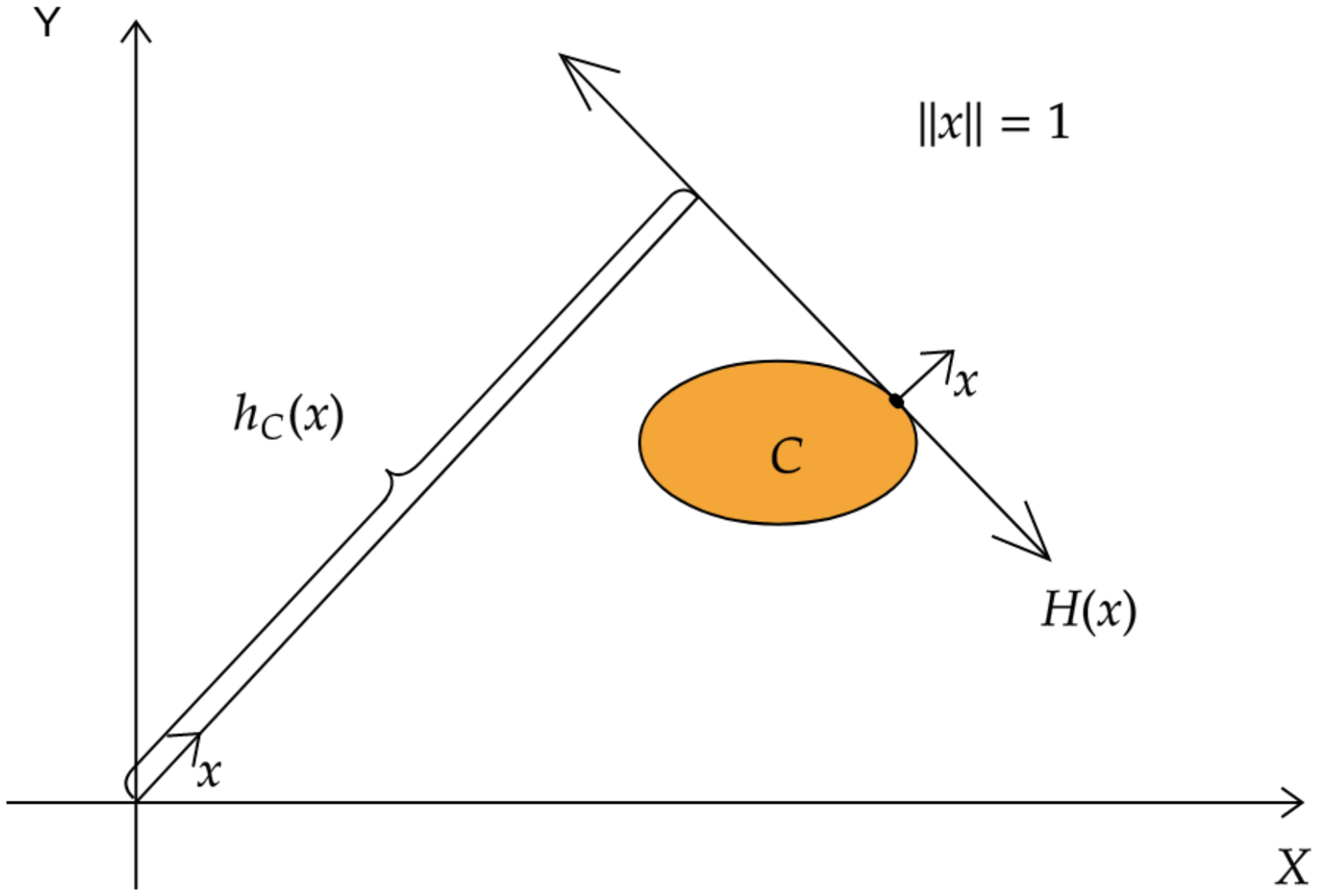}
    {Graphical representation of the support function and supporting hyperplane. }
\end{center}

\end{mdframed}

\begin{mdframed}[style=MyFrame]
\subsection*{Kernel functions}
The origins of the kernel function lie in Reproducible Kernel Hilbert Spaces (RKHS).
Let us consider a Hilbert space $H$ consisting of real-valued functions defined on an arbitrary set $X.$
We can define an evaluation functional $E_x(f)=f(x)$ mapping each function to a real number. If this functional for all $x \in X$ is bounded at each $f \in H,$  we call $H$ an RKHS. Applying Riesz's theorem mentioned in the introduction, we have the following  representation $E_x(f)= \braket{f}{K_x}$
where $K_x\in H.$
 It follows that $E_y(K_x)=\braket{K_x}{K_y}=K_x(y)=K(x,y).$

The function $K(x,y) :X\times X \to \mathbb{R}$ is called the Reproducing Kernel. Succinctly, in an RKHS the evaluation of a function at a point can be replaced by taking an inner product with a function determined by the kernel in the associated function space.

Now given a feature map $\Phi:X\to \Gamma$ where $\Gamma$ is a Hilbert space, we can define a normed space
$
H_{\varPhi }=\{f:X\to \mathbb {C} \mid \exists g\in \Gamma,f(x)=\braket{g}{\varPhi (x)}\forall {\text{ }}x\in X\}$
with the norm
$\|f\|_{\varPhi }=\inf\{\|g\|_{\Gamma}:g\in G,f(x)= $ $\braket{g}{\varPhi (x)},\forall {\text{ }}x\in X\}.$ It can be proven that ${H_{\varPhi }}$ is a RKHS and its kernel defined by $K(x,y)= \braket{\varPhi(x)}{\varPhi(y)}$.  
\end{mdframed}

Formally a kernel $k(\mathbf{x}_j, \mathbf{x}_k)$ is defined as a function that computes the inner product of the images of $\mathbf{x}_j, \mathbf{x}_k$ in the feature space. That is $k(\mathbf{x}_j, \mathbf{x}_k) = \Lambda(\mathbf{x}_j)^\top \Lambda(\mathbf{x}_k)$. By the virtue of the equality established here all the inner products in \eqref{eq:hyperplane_svd}, \eqref{eq:hyperplane_svd_b} and \eqref{eq:alpha_opt} can be replaced by the corresponding kernels. We can then use the kernel trick: that is to assign an analytic expression for the kernel in \eqref{eq:hyperplane_svd}, \eqref{eq:hyperplane_svd_b} and \eqref{eq:alpha_opt}, provided that the expression satisfies all the required conditions for it to be a valid kernel \cite{schol02}. Indeed every choice of a kernel can be implicitly associated to a feature map $\Lambda$. However, in the current approach we do not need to know the explicit form of the feature map. In fact this is the advantage of the kernel trick as calculating $\Lambda({\bf x})$ is less efficient than using the kernels directly.

Most modern applications in machine learning however involves deep neural networks (DNN) \cite{sch12}.  A DNN can also be looked upon as a collection of perceptrons arranged in a definite fashion. The architecture of a DNN is best understood and visualised in the form of a graph. A DNN is composed of an input layer, an output layer and several hidden layers in between. Each layer comprises of several nodes. Each of these nodes represent a neuron. Edges are allowed to exist between nodes belonging to nearest neighbouring layers only. That is nodes in layer $j$ share edges with nodes in layer $(j+1)$ and those in layer $(j-1)$. For the sake of simplicity we consider the case where all nodes in the $j$-th layer are connected to all nodes in the $(j\pm1)$-th layer---a fully connected neural network.

A DNN takes a data ${\bf x}$ as an input (at the input layer). This input vector is subsequently manipulated by the neurons in the next layer (the first hidden layer) to output a transformed vector ${\bf x}^{(1)}$, and this process repeats until the last layer (output layer) is reached. Consider the $k$-th neuron in the $j$-th layer. For convenience we would denote this neuron as $(k,j)$. It receives an input vector ${\bf x}^{(j-1)}$, whose components are the outputs of the neurons in the $(j-1)$-th layer. The vector is then transformed by the $(k,j)$ neuron as:

\begin{equation}
\label{eq:dnn_comp}
    {\bf x}^{(j-1)} \rightarrow \Psi\Big(({\bf w}^{(j-1)}_k)^\top\cdot{\bf x}^{(j-1)} + b^{(j)}_k\Big) = \mathbf{x}^{(j)}_k
\end{equation}
where ${\bf w}^{(j-1)}_k$ are the weights associated with the edges that connect the neuron $(k,j)$ to the neurons in the previous layer, $b^{(j)}_k$ is the bias corresponding to the neuron $(k,j)$ and $\Psi(\cdot)$ is a differentiable non-linear function also known as the activation function. This is the fundamental mathematical operation that propagates the data in a DNN, layer by layer, in the forward direction (input layer to output layer), through a series of complicated transformations. The training component of the algorithm is however accomplished through the back-propagation  step \cite{roj96} ---a cost function is calculated by comparing the signal at the output layer (the model predicted label for the data ${\bf x}$) and the desired signal (the actual label for the data ${\bf x}$), based on which the weights and the biases are adjusted such that the cost function is minimised. Apart from supervised learning DNN are routinely used for unsupervised learning including generative learning. In generative learning, given access to a training dataset, a machine learning model learns its underlying probability distribution for future sample generation. To formulate this mathematically, consider a dataset $\mathcal{S} =\{\mathbf{x}_j\}$, whose entries $\mathbf{x}_j \in \mathbb{R}^N$ are independent and identically distributed vectors and are sampled as per a distribution $q({\mathbf{x}})$. The purpose of a generative model is to approximate $q({\mathbf{x}})$, given the access to training data from the dataset $\mathcal{S}$. To achieve this, a machine learning model (with  tunable parameters $\boldsymbol{\theta}$) is trained such that the model generated distribution, $p({\mathbf{x}}, {\boldsymbol{\theta}})$, mimics the true distribution. The standard practice in generative learning is to minimize the negative log-likelihood with respect to the model parameters which tantamount to minimizing the Kullback-Leibler divergence $D_{KL}(q({\bf x}) \vert \vert p({\bf x}, \boldsymbol{\theta}))$ between the two distributions.

\subsection{Variational algorithms and quantum machine learning}
Variational quantum computing has emerged as the preeminent model for quantum computation. The model merges ideas from machine learning to better utilise modern quantum hardware.

Mathematically the problem in variational quantum computing can be formulated as---given (i) a variational quantum circuit (a.k.a.~ansatz) $U(\boldsymbol{\theta}) \in \text{U}_\mathbb{C}(2^n)$ which produces a $n$-qubit variational state $\ket{\psi(\boldsymbol{\theta})} = U(\boldsymbol{\theta}) \ket{0}^{\otimes n}$; $\boldsymbol{\theta} \in [0, 2\pi)^{\times p}$ (ii) an objective function $\mathcal{H} \in \text{herm}_{\mathbb{C}} (2^n)$ and (iii) the expectation $\bra{\psi(\boldsymbol{\theta})} \mathcal{H} \ket{\psi(\boldsymbol{\theta})}$ find

\begin{equation}
\boldsymbol{\theta}^\star \in \underset{{\boldsymbol{\theta} \in [0, 2\pi)^{\times p}}} {\arg\min} \bra{\psi({\boldsymbol{\theta}})} \mathcal{H} \ket{\psi({\boldsymbol{\theta}})}.
\end{equation}
$\ket{\psi (\boldsymbol{\theta}^\star)}$ in this case would approximate the ground state (eigenvector corresponding to the lowest eigenvalue) of the Hamiltonian $\mathcal{H}.$ $\mathcal{H}$, often called the problem Hamiltonian, can embed several problem classes such that the solution to the problem is encoded in its ground state.  The variational model of quantum computation was shown to be universal in \cite{Bia21}. 

Quantum machine learning both discriminative and generative emerged as an important application of variational algorithms with suitable modifications to the aforementioned scheme. Indeed by their very design variational algorithms are well suited to implement machine learning tasks on a quantum device. The earlier developments in QML came mainly in form of classification tasks \cite{Mit+18,FN18}.

Classification of a classical dataset $\mathcal{S} = \{(\mathbf{x}_j, \mathbf{y}_j))\}_{j=1}^M$ on quantum hardwares typically involves four steps. Firstly the input vector $\mathbf{x}_j$ is embedded into a $n$ qubit state $\ket{\psi(\mathbf{x}_j)}$. The effect of data encoding schemes on the  expressive power of a quantum machine learning model was studied in \cite{SSM21}. What is the most effective data embedding scheme? Although there are a few interesting candidates \cite{Llo+20, Sal+20}, largely this question remains unanswered. In the second step a parameterised ansatz $U(\boldsymbol{\theta})$ is applied on $\ket{\psi(\mathbf{x}_j)}$ to output $\ket{\psi(\mathbf{x}_j, \boldsymbol{\theta})}$. A number of different ansatze are in use today, including the hardware efficient ansatz, the checkerboard ansatz, the tree tensor network ansatz etc., which are chosen as per the application and implementation specifications. The third step in the process is where data is read out of $\ket{\psi(\mathbf{x}_j, \boldsymbol{\theta})}$: expectation values of certain chosen observables (Hermitian operators) are calculated with respect to $\ket{\psi(\mathbf{x}_j, \boldsymbol{\theta})}$ to construct a predicted label $F(\mathbf{x}_j, \boldsymbol{\theta})$. The measured operators are typically the Pauli strings which form a basis in $\text{herm}_\mathbb{C}(2^n).$ In the final step a cost function is constructed as in \eqref{eq:loss_fn_formal} and minimised by tuning $\boldsymbol{\theta}$. This approach has been used in several studies to show successful classifications in practical datasets (see e.g.~\cite{Sch+20}).

An interesting variation of the aforementioned approach was shown in \cite{SK19,Hav+19} to implement data classification based on the kernel trick. In this method the Hilbert space is treated as a feature space and the data embedding step---$\mathbf{x}_j \rightarrow \ket{\psi(\mathbf{x}_j)}$---as a feature map. The quantum circuit is used directly to compute the inner product $\braket{\psi(\mathbf{x}_j)}{\psi(\mathbf{x}_k)}$, using e.g.~the swap test, which is then used for data classification using classical algorithms such as SVMs. 

Quantum machine learning has also been used to classify genuine quantum data. Some prominent examples of such applications include: classifying phases of matter \cite{UKB20}, quantum channel discrimination \cite{Kar+20} and entanglement classification \cite{Gra+18}. Other machine learning problems with quantum mechanical origins that has been solved by variational algorithms include e.g.~quantum data compression \cite{ROG17} and denoising of quantum data \cite{BF20}. Both of these applications use a quantum autoencoder. A quantum autoencoder much like their classical counterparts consists of two parts--- an encoder and a decoder. The encoder removes the redundant information from the input data to produce a minimal low dimensional representation. This process is known as feature extraction. To ensure that the minimal representation produced by the encoder is efficient, a decoder is used which takes the output of the encoder and tries to reconstruct the original data. Thus in an autoencoder both the encoder and the decoder are trained in tandem to ensure that the input at the encoder and the output at the decoder closely match each other. While in the classical case the encoders and the decoders are chosen to be neural networks, in the quantum version of an autoencoder neural networks are replaced by variational circuits.

Much advances were made in the front of quantum generative learning as well. In \cite{Ben+19} it was shown that generative modelling can be used to prepare quantum states by training shallow quantum circuits. The central idea is to obtain the model generated probability distribution $p(\boldsymbol{\theta})$ by making repeated measurements on a variational state $\ket{\psi(\boldsymbol{\theta})}$. The state $\ket{\psi(\boldsymbol{\theta})}$ is prepared on a short depth circuit with a fixed ansatz and parameterized with the vector $\boldsymbol{\theta}$. The target distribution $q$ is also constructed similarly, by making repeated measurements on the target state. The measurement basis (preferably informationally complete positive operator valued measures), as expected, is kept to be the same in both the cases. The training objective therefore is to ensure that $p(\boldsymbol{\theta})$ mimics $q$ such that the variational circuit learns to prepare the target state. The same task in an alternate version can be looked upon as a machine learning assisted quantum state tomography \cite{car+19}.

\section{Tensor networks in machine learning}

\subsection{Tensor networks in classical machine learning}

Recently tensor network methods have found several applications in machine learning. Here we discuss some of these applications with a focus on supervised learning models. We return to our labelled dataset $\mathcal{S} = \{(\mathbf{x}_j, \mathbf{y}_j)\}_{j=1}^M$, where $\mathbf{x}_j \in \mathbb{R}^N$. As mentioned earlier there are several machine learning models $F$ to choose from, to perform a classification on the dataset $\mathcal{S}$.

However more abstractly speaking, a classifier $F$ can be expressed as the following function:

\begin{equation}
\label{eq:model_poly}
F_{\bf W}(\mathbf{x}) = \sum_{j_1,j_2,\cdots, j_N \in \{0,1\}} W_{j_1 j_2\cdots j_N}x_1^{j_1}x_2^{j_2}\cdots x_N^{j_N},
\end{equation}
in the polynomial basis \cite{nov16}. Here $\mathbf{x}\in \mathbb{R}^N $ is an input datum and $x_k\in \mathbb{R}$ is the $k$-th component of $\mathbf{x}$. The tensor $W_{j_1 j_2\cdots j_N}$ is what we will call as the weight tensor, which are the tunable parameters in $F$. Going back to the case of binary classification, that is $\mathbf{y}_j \in \{1, -1\}$, $F(\mathbf{x})$ can be looked upon as a surface in $\mathbb{R}^{N+1}$ that can be tuned (trained) such that it acts as a decision boundary between the two classes of input data. Indeed the training can be accomplished by the minimisation:

\begin{equation}
    \min_{\bf W} \Big( \sum_{j=1}^M \vert sgn(F_{\bf W}(\mathbf{x}_j)) - \mathbf{y}_j) \vert^2 \Big),
\end{equation}
where $sgn(F_{\mathbf{W}}(\mathbf{x}_j))$ is the predicted label. 

However in practice we encounter a bottleneck while computing \eqref{eq:model_poly} as it involves $2^N$ components of the weight tensor. One way to circumvent this bottleneck is to express the weight tensor as a MPS. Following the observation in section \ref{sec:ten_net}, for a suitable choice of the virtual bond dimension $r$, the MPS representation of the weight tensor $W$ would involve only $O(\text{poly}~N)$ components thus making the computation of $\eqref{eq:model_poly}$ less resource intensive \cite{SS16}. Here it is prudent to note that we could alternatively have opted for any other basis in the decomposition of the function $F$ depending on the optimization problem at hand.

Yet another application of tensor networks in machine learning can be seen in DNN. Consider the transformation in \eqref{eq:dnn_comp}. For most practically relevant DNN this transformation is highly resource intensive. This is due to the fact that the vectors $\mathbf{x}^{(j-1)}$ are typically very large and hence computing the inner product $({\bf w}^{(j-1)}_k)^\top\cdot{\bf x}^{(j-1)}$ is difficult. This computation can be made efficient however by using MPS. In order to do this the vectors ${\bf w}^{(j-1)}_k, {\bf x}^{(j-1)}$ are first reshaped thus converting them into tensors and then expressed as a MPS. Upon expressing a vector as a MPS we would need to keep track of much fewer components compared to the original representation. This makes the computation of \eqref{eq:dnn_comp} tractable.

\subsection{The parent Hamiltonian problem}

Consider the quantum state preparation problem using a variational algorithm. Given a variational circuit $U(\boldsymbol{\theta})$ and an $n$ qubit target state $\ket{t}$, tune $\boldsymbol{\theta} \rightarrow \boldsymbol{\theta}^\star$ such that $U(\boldsymbol{\theta}^\star) \ket{0}^{\otimes n}$ approximates $\ket{t}$. To accomplish this task one needs to construct a Hamiltonian $\mathcal{H} \in \text{herm}_{\mathbb{C}} (2^n)$ with $\ket{t}$ as its unique ground state, which would serve as the objective function of the algorithm. Constructing such a Hamiltonian for a given target state is known as the parent Hamiltonian problem. The simplest construction of such a Hamiltonian is $\mathcal{H} = \eye - \ket{t}\bra{t}$. This construction is however not always useful as expressing $\mathcal{H}$ in the basis of Pauli strings--the basis of measurement-- may require an exponential number of terms. Thus estimating the expectation of $\mathcal{H}$ in polynomial time becomes impossible. 

Ideally we want the Hamiltonian to have the following properties
\begin{enumerate}
    \item The Hamiltonian is non-negative.

    \item The Hamiltonian must have a non-degenerate (unique) ground state $\ket{t}$.
    
    \item The Hamiltonian must be gapped. A $n$ qubit Hamiltonian $\mathcal{H}(n) \geq 0$ is said to be gapped if 
    \begin{equation} \label{eq:gap}
           \lim_{n \rightarrow \infty} \Big[ \dim \ker{\mathcal{H}(n)} \Big] = 1.
    \end{equation}
    Validity of \eqref{eq:gap} ensures that $\mathcal{H}(n)$ is gapped for all finite $n$.
    
    \item The Hamiltonian must be local. A $n$ qubit Hamiltonian $\mathcal{H}(n)$ is said to be local if $\mathcal{H}(n)$ can be expressed as
    
    \begin{equation}
        \mathcal{H}(n) = \sum_{j \in 2^V} h(j)
    \end{equation}
    where $V$ is the set of $n$ symbols (qubits) and $h(j) = \bigotimes_{k \in j} P_k \in \text{herm}_{\mathbb{C}} (2^n)$, where $P \in \text{herm}_{\mathbb{C}}(2)$. $\mathcal{H}(n)$ will be called $k-$local if $\nexists h(j)$ that operates on more than $k$ symbols (qubits) non-trivially. Here a trivial operation refers to the case when $P_k$ identity for a certain $k$.

    \item The Hamiltonian must have $O(\text{poly}~n)$ terms when expressed in the Pauli basis. The number of terms in a Hamiltonian when expressed in the Pauli basis is also known as the cardinality of the Hamiltonian, $|| \mathcal{H} ||_{card}$ \cite{Bia21}.
\end{enumerate}

Hamitonians with such properties can indeed be constructed if $\ket{t}$ admits a matrix product state, albeit with an additional condition that $\ket{t}$ must satisfy the condition of injectivity. For the parent Hamiltonian construction consider the following setting. Let $\ket{t}$ be a $n$ qubit state written as a translation invariant and site independent MPS with periodic boundary conditions
\begin{equation}
    \ket{t} = \sum_{j_1 \cdots j_n} Tr(A_{j_1} \cdots A_{j_n}) \ket{j_1 \cdots j_n}
\end{equation}
where $A_{j_k} \in \text{Mat}_{\mathbb{C}}(r)$. For the sake of brevity we will call these matrices as Kraus operators.\footnote{Indeed there is a connection between the matrices $A_{j_k}$ and completely positive trace preserving (CPTP) maps from which $A_{j_k}$ derives their name. For the purpose of this paper however we will skip the detailed discussion.} Consider the map
\begin{equation}
    \Gamma_L : X \rightarrow \ket{\psi^{(L)}}_X = \sum_{j_1...j_L}Tr(XA_{j_1} \cdots A_{j_L}) \ket{j_1 j_2 ... j_L}
\end{equation}
where $X \in \text{Mat}_{\mathbb{C}}(r)$. We say that the state $\ket{t}$ is injective with injectivity length $L$ if $\Gamma_L$ is an injective map. Several corollaries follow from this definition. A particularly useful one connects the notion of injectivity to rank of reduced density matrices. It asserts that for a $L$-qubit reduced density matrix---$\rho^{(L)}$---of $\ket{t}$,  $\text{rank}(\rho^L) = r^2$ if injectivity holds. It has been shown that in the large $n$ limit, $\rho^{(L)}$ is given by 
\begin{equation} \label{eq:red_den}
    \rho^{(L)} = Tr_{n-L}(\ketbra{t}{t}) = \sum_{\alpha, \beta = 1}^r \Lambda_\alpha \ketbra{\psi_{\alpha \beta}^{(L)}}{\psi_{\alpha \beta}^{(L)}}
\end{equation}
with $\ket{\psi_{\alpha \beta}^{(L)}} = \sum_{j_1 \cdots j_L} \braket{\alpha}{A_{j_1} \cdots A_{j_L}|\beta} \ket{j_1 \cdots j_L}$, $\ket{\alpha}, \ket{\beta} \in \mathbb{C}^r$ and $\Lambda_\alpha \in \mathbb{R}_+$. Alternately this would mean that $\{ \ket{\psi_{\alpha \beta}^{(L)}} \}_{\alpha \beta}$ is a linearly independent set.

The from of the reduced density matrix in \eqref{eq:red_den} is particularly telling and allows us to construct the parent Hamiltonian of $\ket{t}$: $\mathcal{H} \geq 0$. We formally write our parent Hamiltonian as

\begin{equation} \label{eq:ph}
    \mathcal{H} = \sum_{j=1}^n h_j^{(L)}
\end{equation}
where $h_j^{(L)}$ operates non-trivially over $L$ qubits from $j$ to $L+j$ and obeys the condition $\ker h_j^{(L)} = \Span{\{ \ket{\psi_{\alpha \beta}^{(L)}}} \}_{\alpha \beta}$. The latter condition combined with \eqref{eq:red_den} ensures that $\text{Tr}(h_j^{(L)} \rho_j^{(L)}) = 0$ for all $j$, which in turn implies that $\ket{t} \in \ker \mathcal{H}$. In fact $\ket{t} = \ker \mathcal{H}$, provided $\ket{t}$ is injective thus satisfying condition-1 for $\mathcal{H}$. Conditions 3 and 4 are satisfied naturally due to the form of $\mathcal{H}$ in \eqref{eq:ph}. In addition $\mathcal{H}$ can also seen to be frustration free, that is $\braket{t}{\mathcal{H}|t} = 0 \implies \braket{t}{h_j^{(L)}|t} = 0 ~\forall j$. Finally it was shown in \cite{FNW92} that if $\ket{t}$ is injective, $\mathcal{H}$ is gapped.

\section{Conclusion}

The importance of matrix product states in physics is due to the ease at which one might calculate important properties, such as two-point functions, thermal properties and more. This is also true in machine learning applications.  For example, images of size 256x256 can be viewed as rank one tensor networks on $\mathbb{R}^{256}$. Departing from this linear (train) structure results in tensors with potentially much greater expressability at the cost of many desirable properties being lost.

\printnotes

 \bibliography{bibliography.bib}
 \bibliographystyle{plain}
 
\begin{info}
Richik Sengupta [\url{r.sengupta@skoltech.ru}] is a research scientist at Skolkovo Institute of Science and Technology.\\

Soumik Adhikary [\url{s.adhikari@skoltech.ru}] is a research scientist at Skolkovo Institute of Science and Technology.\\

Ivan Oseledets  [\url{i.oseledets@skoltech.ru}] is Full Professor, Director of the Center for Artificial Intelligence Technology, Head of the Laboratory of Computational Intelligence at Skolkovo Institute of Science and Technology.\\

Jacob Biamonte [\url{j.biamonte@skoltech.ru}] is Full Professor, Head of the Laboratory of Quantum algorithms for machine learning and optimisation at Skolkovo Institute of Science and Technology.   
\end{info}


\end{document}